\documentclass[aps,twocolumn]{revtex4-1}

\usepackage{slashed}
\usepackage{stmaryrd}
\usepackage{color}
\usepackage{mathrsfs}
\usepackage{yfonts}
\usepackage{rotating}
\usepackage{graphicx}
\usepackage{amsfonts}
\usepackage{amsmath}
\usepackage{mathrsfs}
\usepackage[american]{babel}
\usepackage{wasysym}
\usepackage{slashed}
\usepackage{pgfplots}
\usepackage{braket}
\usepackage{caption}
\usepackage{graphicx}
\usepackage{subfigure}

\newcommand{\g}[0]{\gamma}
\newcommand{\al}[0]{\alpha}
\newcommand{\be}[0]{\beta}
\newcommand{\de}[0]{\delta}
\newcommand{\e}[0]{\varepsilon}

\newcommand{\si}[0]{\sigma}

\begin{document}
\title{Quantum vacuum polarization around a Reissner-Nordstr\"om black
  hole in five dimensions}
\author{Gon\c{c}alo M. Quinta}
\email{goncalo.quinta@ist.utl.pt}
\affiliation{Centro de Astrof\'{\i}sica e Gravita\c c\~ao - CENTRA,
Departamento de F\'{\i}sica, Instituto Superior T\'ecnico - IST,
Universidade de Lisboa - UL, Avenida Rovisco Pais 1,
1049-001 Lisboa, Portugal}
\author{Antonino Flachi}
\email{flachi@phys-h.keio.ac.jp}
\affiliation{Department of Physics, \& Research and Education Center for Natural Sciences, Keio University, Hiyoshi 4-1-1, Yokohama, Kanagawa 223-8521, Japan}
\author{Jos\'{e} P. S. Lemos}
\email{joselemos@ist.utl.pt}
\affiliation{Centro de Astrof\'{\i}sica e Gravita\c c\~ao - CENTRA,
Departamento de F\'{\i}sica, Instituto Superior T\'ecnico - IST,
Universidade de Lisboa - UL, Avenida Rovisco Pais 1,
1049-001 Lisboa, Portugal, and \\
Gravitational Physics Group,
Faculty of Physics, University of Vienna, Boltzmanngasse 5,
A1090, Wien, Austria}

\begin{abstract}
WKB approximation methods are applied to the case of a massive scalar field around a five-dimensional Reissner-Nordstr\"om black hole. The divergences are explicitly isolated and the cancellation against the Schwinger-DeWitt counterterms are proven. The resulting finite quantity is evaluated for different values of the free parameters, namely the black hole mass and charge, and the scalar field mass.
We thus extend our previous results on quantum vacuum polarization effects for uncharged
asymptotically flat higher-dimensional black holes to  electrically charged  black holes.
\end{abstract}
\maketitle

\newpage

\section{Introduction}

In \cite{Flachi:2016bwr} we have adapted the WKB method to
higher-dimensional Schwarzschild black holes and explicitly calculated
the scalar vacuum polarization for the case of five, i.e., (4+1),
dimensions everywhere outside the horizon.  Other works that have
studied vacuum polarization in higher-dimensions are
\cite{Frolov:1989rv} in which the five-dimensional case is also
treated,~\cite{Shiraishi:1993ti} in which odd-dimensional anti-de
Sitter spacetime is considered,~\cite{Decanini:2005eg,Thompson:2008bk}
in which renormalization in higher-dimensions is developed with care,
\cite{Breen:2015hwa} which studied vacuum polarization on branes,
and~\cite{Breen:2016, Breen:2017} in which scalar vacuum polarization,
using a mode-sum regularization prescription, is computed for
higher-dimensional Schwarzschild black holes with explicit results up to
11 dimensions.  The initial studies in vacuum polarization in curved
spacetimes~\cite{christensen,Candelas:1980zt,Candelas:1984pg,Fawcett:1983dk,Anderson:1989vg}
focused in four, i.e., (3+1), dimensions and had the aim of improving
the understanding of particle production in curved spacetimes and
various aspects of black hole evaporation.

Following~\cite{Flachi:2016bwr}, in which
higher-dimensional Schwarz\-schild black holes were studied, here we
adapt again the WKB method originally devised
in~\cite{Candelas:1980zt,Candelas:1984pg} to the case of higher-dimensional
Reissner-Nordstr\"om black holes.

The paper is organized as follows. In Sec.~\ref{vacP}, we will outline the standard properties of the Green function and its mode-sum decomposition in a five-dimensional spacetime. In Sec.~\ref{WKBap}, the WKB method is used to obtain a truncated approximation of the Green function. In Sec.~\ref{Renorm}, we use the point-splitting method to renormalize the coincidence limit of the Green function, i.e. the vacuum polarization, regularizing first the summation in the angular modes followed by the energy modes. In Sec.~\ref{Num}, we numerically compute the previously calculated renormalized vacuum polarization, providing results for different values of black hole mass, charge and scalar field mass. In Sec.~\ref{Conc}, we draw some conclusions.

\section{Vacuum Polarization in Higher-Dimensions}
\label{vacP}

We are interested in the vacuum polarization $\braket{\phi^2(x)}$ of a scalar quantum field, which is given by the coincidence limit of the associated Euclidean Green function
$G_\textrm{E}$, which satisfies the differential equation
\begin{equation}
  \left(\square_{\textrm{E}} - \mu^2 - \xi R\right)G_\textrm{E}(x,x') = - {\de(x-x') \over \sqrt{|g|}}\,,
\label{gf}
\end{equation}
where $\square_\textrm{E}$ is the d'Alembertian operator with Euclidean signature, $\mu$ is the scalar field mass, $\xi$ is the coupling constant, $R$ is the spacetime curvature, and
$x$ and $x'$ are spacetime points.

In this work, we will consider the background to be a five-dimensional black hole
described by a five-dimensional metric of the type
\begin{equation}\label{RS5}
ds^2 = -f(r) dt^2 + {dr^2 \over f(r)} + r^2 d\Omega^2_3\,,
\end{equation}
where $t$ and $r$ are the time and radial coordinates, respectively, $d\Omega^2_3$ represents the line element of a 3-sphere, and
$f(r)$ is some function of $r$. We assume that at infinity
$f(r)$ goes as $1/r^2$ as it should for
a five-dimensional spherical asympotically flat spacetime,
and we also assume that $f(r)$ contains a horizon
at some radius $r_+$.

Performing a Wick rotation $t = -i \tau$ on the time coordinate, we obtain the Euclidean metric
\begin{equation}
ds_{\rm E}^2 = f(r) d\tau^2 + {dr^2 \over f(r)} + r^2 d\Omega^2_3\,,
\end{equation}
which is positive definite everywhere outside the horizon. In order to avoid conical singularities in the Euclidean metric, the coordinate $\tau$ must be periodic with period $\be$ equal to
\begin{equation}
\be = 4\pi \left({df \over dr}\right)^{-1}_{r = r_+}\,.
\label{temp}
\end{equation}
The quantity $T = \be^{-1}$ will then be the characteristic temperature of the black hole.

Working in the Hartle-Hawking vacuum state, we may write the finite temperature Euclidean Green function in the mode-sum representation
\begin{equation}
  G_{\rm E}(x,x') = {\al \over 4\pi^3} \sum^{\infty}_{n=-\infty} e^{i \omega_n \Delta \tau} \sum^{\infty}_{l=0} (l+1) C^{(1)}_l(\cos \g) G_{nl}(r,r')
  \label{ms}
\end{equation}
where $\al \equiv 2\pi / \be$,
$\Delta \tau = \tau - \tau'$, $\omega_n \equiv \al n$, $\g$ is the geodesic distance in the 3-sphere,
and $C^{(1)}_l(x)$ is a Gengenbauer polynomial.
Inserting
the mode-sum expansion, Eq.~(\ref{ms}), in Eq.~(\ref{gf})
leads to the differential equation for the radial Green function
\begin{align}\label{rgf}
\bigg\{{d \over dr}&\left(r^{3}f(r){d \over dr}\right) - r^3 \left({\omega^2_n \over f(r)} + \mu^2 +\xi R \right) \nonumber \\
& \hspace{10mm} - l(l+2)r \bigg\}G_{nl}(r,r') = - \de(r-r')\,.
\end{align}
The solution to Eq.~(\ref{rgf}) can be expressed in terms of solutions of the corresponding homogeneous equation. In particular, if $p_{nl}(r)$ and $q_{nl}(r)$ are solutions of the homogeneous equation regular at the horizon and infinity, respectively, then the radial Green function can be written as
\begin{equation}\label{rgfrr}
G_{nl}(r,r') = C_{nl} \, p_{nl}(r_<) q_{nl}(r_>)
\end{equation}
where $r_<$ and $r_>$ denote the largest and the smallest values of the set $\{r,r'\}$. The quantity $C_{nl}$ is a normalization constant, given by
\begin{equation}
C_{nl} = - {1 \over r^3 f(r)} {1 \over \mathcal{W}(p_{nl}(r),q_{nl}(r))}
\end{equation}
where ${\mathcal W}(p,q)$ is the Wronskian of the two functions.

We now want to find the solution of Eq.~(\ref{rgf}). We will first present the approximate
limiting solutions at infinity and at the horizon and then we develop
the general solution. The limiting solutions serve as boundary conditions
for the general solution. In particular they are useful for
numerical calculations checking.

\section{WKB approximation}
\label{WKBap}

\subsection{Near-infinity and near-horizon solutions}

The form of $p_{nl}$ and $q_{nl}$ of the Green function in Eq.~(\ref{rgfrr}),
solution of Eq.~(\ref{rgf}),
can be obtained by expressing the homogeneous equation in two
limits, namely, the near-infinity limit and the
near-horizon limit.

Starting with the near-infinity limit, i.e.,
the large $r$ limit, the homogeneous equation of Eq.~(\ref{rgf}) becomes
\begin{equation}
\left\{{d^2 \over dr^2} + {3 \over r} {d \over dr} - (\omega^2_n + \mu^2 + \xi R)\right\}q_{nl}(r) = 0\,,
\end{equation}
the solution of which, regular at infinity, is of the form
\begin{equation}\label{qsol}
q_{nl}(r) \sim r^{-3/2} e^{- r \sqrt{\omega^2_n + \mu^2 + \xi R}}\,.
\end{equation}

The near-horizon limit may be obtained by using the tortoise coordinate $r_*$, defined through
$
dr_* = {dr \over f(r)}
$,
in terms of which, in the near-horizon limit and for $n \neq 0$, the homogeneous equation of Eq.~(\ref{rgf}) becomes
\begin{equation}\label{nn0}
\left({d^2 \over dr^2_*} - \omega^2_n\right) p_{nl}(r)=0\,.
\end{equation}
The solution of Eq.~(\ref{nn0}), regular at the horizon, is given by
\begin{equation}\label{psol}
p_{nl}(r) \sim {e^{-\omega_n r_*} \over r}\,.
\end{equation}
In the case $n = 0$, the homogeneous equation of Eq.~(\ref{rgf}), in the near-horizon limit, becomes
$
  {d \over dr}(\ln p_{0l}(r)) = {1 \over f'(r)}\left( {l(l+2) \over r^2} + \mu^2 + \xi R \right)
$,
the solution of which
goes as
\begin{equation}\label{psol0}
p_{0l}(r) \sim \exp\left\{ \int^{r}_{r_+} \left( {l(l+2) \over u^2} + \mu^2 + \xi R \right) {du \over f'(u)} \right\}\,.
\end{equation}

These limiting solutions will be especially important when performing numerical computations, since they will provide the boundary conditions necessary to solve Eq.~(\ref{rgf}) numerically.

\subsection{WKB general solution}

We shall now display a general solution of Eq.~(\ref{rgf}) by following the standard procedure developed in \cite{Candelas:1980zt,Candelas:1984pg}, which makes use of a WKB approximation. We begin by using the following ansatz for the solutions of the homogeneous equation for the radial Green function
\begin{align}
p_{nl}(r) & = {1 \over \sqrt{r^{3} W(r)}} \exp \left\{ + \int^{r}_{r_+} {W(u) \over f(u)} \,  du\right\} \label{psolWKB}\,, \\
q_{nl}(r) & = {1 \over \sqrt{r^{3} W(r)}} \exp \left\{ - \int^{r}_{r_+} {W(u) \over f(u)} \, du\right\} \label{qsolWKB}\,,
\end{align}
where $W$ is the WKB function to be determined.
The above expressions are chosen specifically to eliminate all sign dependent terms once inserted in the homogeneous equation of Eq.~(\ref{rgf}), while at the same time satisfying both the near-horizon and large $r$ limits which are going to be calculated below. We will omit the $n$ and $l$ indices in the WKB function $W(r)$ whenever necessary for notational convenience. In the end, we are left with the homogeneous equation
\begin{equation}\label{WKBeq}
W^2 = \Phi + a_1 {W' \over W} + a_2 {W'^2 \over W^2} + a_3 {W'' \over W}\,,
\end{equation}
where
\begin{align}
\Phi & = \left((l+1)^2-1\right){f \over r^2} + \si(r)\,, \\
\si & = \omega^2_n +(\mu^2 + \xi R)f + {3f^2 \over 4 r^2} + {3f f' \over 2r}\,,
\end{align}
and
\begin{equation}
a_1 = {f f' \over 2}, \quad a_2 = -{3 \over 4} f^2, \quad a_3 = {f^2 \over 2}\,,
\end{equation}
where a prime in the functions $W$ and $f$ denotes a derivative with respect to the coordinate $r$.
Inserting Eqs.~(\ref{psolWKB}) and (\ref{qsolWKB}) in Eq.~(\ref{rgfrr}), taking the radial coincidence limit and using the fact the Wronskian is given by $\mathcal{W}(p(r),q(r)) = -f/(2W)$, we obtain
\begin{equation}
G_{nl}(r,r) = {1 \over 2 r^3 W_{nl}(r)}
\end{equation}
The solution to Eq.~(\ref{WKBeq}) can now be expressed iteratively as $W = W_0 + W_1 + \cdots$. At zeroth order, for example, we have $W_0 = \sqrt{\Phi}$. The expansion we are interested in is
\begin{equation}
{1 \over W} = {1 \over \sqrt{\Phi}}(1+\de \Phi + \de^2 \Phi + \ldots)
\end{equation}
were $\de^{n}\Phi / \sqrt{\Phi}$ represents the $n$th order WKB correction to $1/W$. For renormalization purposes, we may only be concerned with the first order approximation, for which one can check that
\begin{equation}
\de \Phi = -{a_1 \over 4} {\Phi' \over \Phi^2} +\left({a_3-a_2 \over 8}\right){\Phi'^2 \over \Phi^3} - {a_3 \over 4} {\Phi'' \over \Phi^2} \,.
\end{equation}
We thus obtain the approximated solution $\tilde{W}$ truncated at first order,
\begin{equation}\label{tildeW}
{1 \over \tilde{W}} = {1+\de \Phi \over \sqrt{\Phi}}\,,
\end{equation}
or, writing explicitly,
\begin{align}\label{tildeW1}
{1 \over \tilde{W}} & = {1 \over \sqrt{\Phi}} + \al_1 {1 \over \Phi^{5/2}} + \al_2 {(l+1)^2 \over \Phi^{5/2}} \nonumber \\
& \hspace{5mm} + \al_3 {1 \over \Phi^{7/2}} + \al_4 {(l+1)^2 \over \Phi^{7/2}} + \al_5 {(l+1)^4 \over \Phi^{7/2}}\,,
\end{align}
with
\begin{align}
\al_1 & = \frac{r \left((a_1 r-4 a_3) f'-a_1 r^3 \sigma
   '+a_3 r f''-a_3 r^3 \sigma ''\right)}{4 r^4} \nonumber \\
   & \hspace{5mm} +\frac{f(6 a_3-2 a_1 r)}{4 r^4}\,, \\
\al_2 & = \frac{f (2 a_1 r-6 a_3)-r \left((a_1 r-4
   a_3) f'+a_3 r f''\right)}{4 r^4}\,, \\
\al_3 & = -\frac{(a_2-a_3) \left(-r f'+2 f+r^3 \sigma
   '\right)^2}{8 r^6}\,, \\
\al_4 & = -\frac{(a_2-a_3) \left(r f'-2 f\right) \left(-r
   f'+2 f+r^3 \sigma'\right)}{4 r^6}\,, \\
\al_5 & = -\frac{(a_2-a_3) \left(r f'-2 f\right)^2}{8 r^6}\,.
\end{align}
Taking the spatial coincidence limit, the Euclidean Green function given in
Eq.~(\ref{ms}) can then be approximated as
\begin{equation}\label{GeDt}
G_{\rm WKB}(x,x') = {\al \over 8\pi^3r^3} \sum^{\infty}_{n=-\infty} e^{i \omega_n \Delta \tau} \sum^{\infty}_{l=0} {(l+1)^2 \over \tilde{W}_{nl}(r)}\,.
\end{equation}

The Euclidean Green function in Eq.~(\ref{GeDt}) is divergent both in the angular and energy modes,
i.e., in the $l$ and $n$ modes, respectively. We take care of this in the following.
The divergence in the angular $l$ modes is purely mathematical and can be promptly removed. On the other hand, the divergent terms in the energy $n$ modes are physical and must be canceled by some counterterms in order to obtain a fully renormalized result.
First we regularize the $l$ modes and afterward the $n$ modes.

\section{Renormalization}
\label{Renorm}

\subsection{Regularization in the $l$ modes}

The summation in the angular modes for large $l$ will be divergent so long as terms of $(l+1)$ with powers larger than $-1$ are present. Expanding $(l+1)^2/\tilde{W}$ for large $(l+1)$, we obtain
\begin{align}
\mathcal{T}_l(r) = & {r \over \sqrt{f}}(l+1) + {r \over 32 f^{3/2} (l+1)}(-16r^2 \omega^2_n+16f \nonumber \\
& \hspace{-5mm} - 4f^2-16f\si + 4rf f'' + r^2 f'^2 - 4r^2f f'')\,,
\end{align}
which diverges in the final sum of Eq.~(\ref{GeDt}). This divergence is not physical, and can be removed by subtracting the quantity
$
{\al \over 8\pi^3r^3} \sum^{\infty}_{n=-\infty} e^{i \omega_n \Delta \tau} \sum^{\infty}_{l=0} \mathcal{T}_l(r)
$,
from Eq.~(\ref{GeDt}). The term involving $\omega^2_n$ is irrelevant, since the summation in $n$ will give $\zeta(-2)$, which is zero. This means the dependence of $\mathcal{T}_l$ is purely on $l$, and so,
$
{\al \over 8\pi^3r^3} \sum^{\infty}_{n=-\infty} e^{i \omega_n \Delta \tau} \sum^{\infty}_{l=0} \mathcal{T}_l(r)
$
is a multiple of $\de( \al \Delta \tau)$. Therefore, since $\Delta \tau \neq 0$, we are effectively subtracting 0, canceling the divergent large $l$ behavior in the process. After the subtraction we may take the full coincidence limit, for which the Green function becomes
\begin{equation}\label{GeDn}
G_{\rm WKB}(x,x) = {\al \over 8\pi^3r^3} \sum^{\infty}_{n=-\infty} \sum^{\infty}_{l=0} \left\{{(l+1)^2 \over \tilde{W}_{nl}} - \mathcal{T}_l\right\}\,.
\end{equation}

\subsection{Regularization in the $n$ modes}

We now proceed to the regularization of the $n$ modes, physically associated to UV divergences. We will isolate the divergent pieces of Eq.~(\ref{GeDn}) and explicitly see that they cancel with the counterterms provided by the point-splitting method developed in \cite{christensen}.

The Green function (\ref{GeDn}) can be written as
\begin{equation}
G_{\rm WKB}(x,x) = {\al \over 8\pi^3r^3} \left(G_0 + 2\sum^{\infty}_{n=1} G_n \right)\,,
\end{equation}
where we have defined $G_n$ as
\begin{equation}\label{Gn}
  G_n = \sum^{\infty}_{l=0} \left(
  {(l+1)^2 \over \tilde{W}_{nl}} - \mathcal{T}_l\right)
\end{equation}
and have made use of the fact that
$\sum^{\infty}_{n=-\infty} G_n =G_0+2\sum^{\infty}_{n=1} G_n $.
The term $G_0$ is finite by construction, so all divergences must be contained within $G_n$. In particular, powers of $n$ larger than $-1$ will result in infinity after the summation. To obtain an expression for $G_n$, we shall make use of the Abel-Plana sum formula
\begin{align}\label{APl}
\sum^{\infty}_{l=j} f(l) & = \int^{\infty}_{0} {dt \over e^{2\pi t} - 1} [f(j+it)-f(j-it)] \nonumber \\
& \hspace{5mm} + {f(j) \over 2} + \int^{\infty}_{j} f(\tau) \, d\tau\,.
\end{align}
Applying Eq.~(\ref{APl}) to Eq.~(\ref{Gn}) and expanding for large $n$, we arrive at the following divergent part of the Green function
\begin{align}\label{Gdiv}
G_{\rm div} & = {\al \over 8 \pi^3 f^{3/2}} \sum^{\infty}_{n=1} \bigg[ \bigg( \mu^2 f - {f \over r^2} + {6\xi f \over r^2} + {f^2 \over r^2} - {6 \xi f^2 \over r^2}  \nonumber \\
& \hspace{-5mm} + {5f f' \over 4r} - {6\xi f f' \over r} - {f'^2 \over 16} + {f f'' \over 4} - \xi f f''\bigg) \ln \omega_n + \omega^2_n \ln \omega_n\bigg]\,.
\end{align}
The divergent terms of the form $1/\omega_n$ cancel out, as expected from spacetimes with odd dimensions; see \cite{Thompson:2008bk}. To obtain a finite renormalized
result we should subtract the counterterms given in
Eq.~(\ref{Gdiv}) from
Eq.~(\ref{GeDn}), i.e.,
\begin{equation}
  G_{\rm reg} = G_{\rm WKB}-G_{\rm div}\,.
 \label{ren1}
\end{equation}

In order to check that Eq.~(\ref{Gdiv}) is the correct
divergent part we use
the generic method devised by Christensen, i.e.,
the  point-slpitting method \cite{christensen}.
Choosing the point split to lie in the $\tau$ coordinate, the geodesic separation
$\sigma$ becomes
\begin{equation}
\si = {f \over 2} \e^2 - {f f'^2\over 96} \e^2 + \mathcal{O}(\e^6)\,,
\end{equation}
and the Schwinger-DeWitt counterterms are then given by
\begin{align}\label{GDS}
G_{\rm SD} & = \lim_{\e \to 0}\bigg\{ {1 \over 16 \pi \sqrt{f} \e} \left(\left( {1\over 6}-\xi \right)R-\mu^2 - {f' \over 4r} + {f'^2 \over 16 f}\right) \nonumber \\
& \hspace{7mm} + {1 \over 8 \pi^2 f^{3/2}\e^3 } \bigg\}\,.
\end{align}
Now, we must express the counterterms as a sum in energy modes, and in order to do that, we convert the inverse powers of $\e$ into sums by using the results
\begin{align}
\lim_{\e \to 0}{1 \over \e} & = - {2 \al \over \pi} \sum^{\infty}_{n=1}\ln \omega_n + \mathcal{O}(\e)\,, \label{e1} \\
\lim_{\e \to 0}{1 \over \e^3} & = {\al \over \pi} \sum^{\infty}_{n=1}\omega^2_n\ln \omega_n + \mathcal{O}(\e)\,, \label{e2}
\end{align}
derived in \cite{Flachi:2016bwr}. Inserting Eqs.~(\ref{e1}) and (\ref{e2}) into Eq.~(\ref{GDS}), one immediately arrives at Eq.~(\ref{Gdiv}), thus confirming its correctness.

\section{Numerical results for the five-dimensional electrically charged
Reissner-Nordstr\"om black hole}
\label{Num}

In obtaining $G_{\rm reg}$ one has made use of the WKB approximation, since
$G_{\rm reg} = G_{\rm WKB}-G_{\rm div}$. We want to go a step further and obtain
a more exact result. The remainder $\de G$ between the
exact value of the Euclidean Green function $G_{\rm E}$
and the WKB approximated Green function $G_{\rm WKB}$, i.e.,
$\de G = G_{\rm E} - G_{\rm WKB}$, is usually ignored because it is considered
negligible. However, here, in our numerical calculation, we take care of
this remainder $\de G$. Thus, instead of writing the approximated
vacuum polarization expression as usual,
$\braket{\phi^2(x)}_{\textrm{ren.}} = G_{\rm reg}$,
we use the exact value for the fully renormalized vacuum polarization as
\begin{equation}
\braket{\phi^2(x)}_{\textrm{ren.}} = G_{\rm reg} + \de G\,.
\end{equation}
The quantity $G_{\rm reg}$ can be evaluated directly using Eq.~(\ref{ren1}).
In the numerical results that follow, we have used the
WKB approximation up to second order and calculated numerically the
remainder $\de G$, which is the most computationally demanding term. In the
process of numerically calculating the remainder, we used
Eqs.~(\ref{psol}) and (\ref{psol0}) for the first point in the
numerical range of the solution (near-horizon limit) and
Eq.~(\ref{qsol}) for the last point (large radius limit).
Of course, if we were to increase
the order of the WKB approximation
in $G_{\rm reg}$, it would reduce the
magnitude of the remainder $\de G$. We have opted to use the
WKB approximation up to second order since it in general
yields accurate results.

In what follows, we specify that the metric given in
Eq.~(\ref{RS5}) is the metric for a five-dimensional electrically charged
Reissner-Nordstr\"om black hole, such that $f(r)$ is given by
\begin{equation}
f(r) = 1 - {2 m \over r^2} + {q^2 \over r^4}\,,
\label{rn2}
\end{equation}
where $m$ is the mass parameter and $q$ is
the electrically charge parameter.
The metric function $f(r)$ given in Eq.~(\ref{rn2}) has an event horizon with radius
\begin{equation}
r_+ = \left( m + \sqrt{m^2 - q^2} \right)^{1/2}\,.
\end{equation}
It has another horizon, the Cauchy horizon, with radius
$r_- = \left( m - \sqrt{m^2 - q^2} \right)^{1/2}$,
but it does not enter into our calculations.
In addition, for the function
$f(r)$ given in Eq.~(\ref{rn2}), the
inverse Hawking temperature defined in Eq.~(\ref{temp})
is
\begin{equation}
\be = {(m + \sqrt{m^2-q^2})^{5/2} \over (m^2 - q^2 + m\sqrt{m^2-q^2})}\pi \,.
\end{equation}
For completeness we remark
that the parameters $m$ and $q$
appearing in Eq.~(\ref{rn2})
are related to the black hole ADM mass
$M$ and electrical charge $Q$, through the relations
$m  = {4 G_5 M \over 3 \pi}$ and
$q^2  = {4 \pi \over 3}G_5 Q^2$, respectively,
where $G_5$ is the gravitational constant for a five-dimensional spacetime.

\begin{figure}
  \centering
    \includegraphics[width=0.5\textwidth]{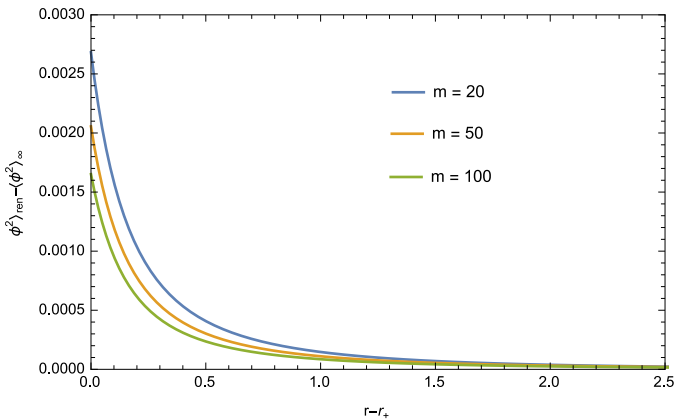}
    \caption{
      Plots of the vacuum polarization
      $\braket{\phi^2}_{\textrm{ren}}-\braket{\phi^2}_{\infty}$ as a
      function of the coordinate distance from the black hole horizon
      radius, i.e., $r-r_+$, for three black hole masses $m$. The charge
      and scalar field mass are fixed as $q = 10$ and $\mu = 0$,
      respectively.}
\end{figure}

\begin{figure}
  \centering
    \includegraphics[width=0.5\textwidth]{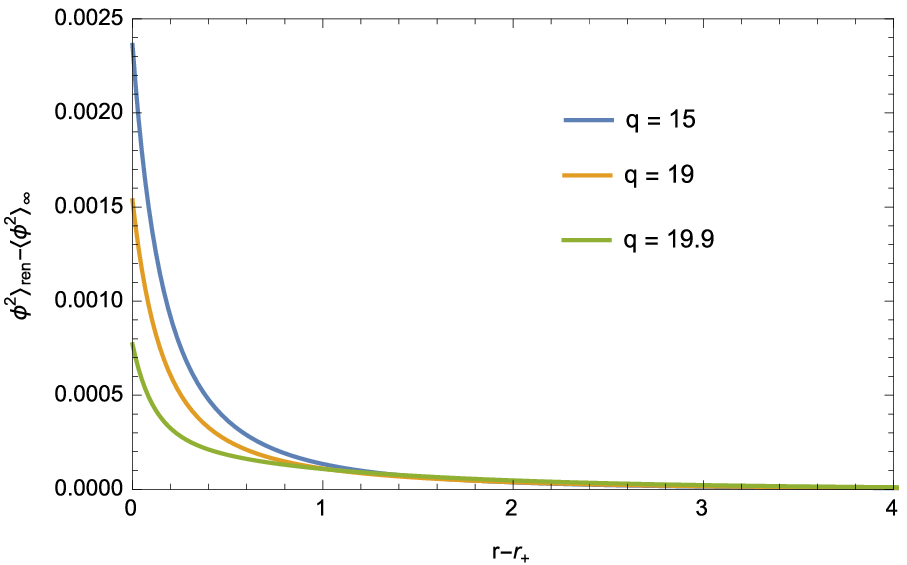}
    \caption{Plots of the vacuum polarization
      $\braket{\phi^2}_{\textrm{ren}}-\braket{\phi^2}_{\infty}$ as a
      function of the coordinate distance from the black hole horizon
      radius, i.e., $r-r_+$, for three black hole charges $q$. The
      back hole and scalar field masses are fixed as $m = 20$ and $\mu
      = 0$, respectively.}
\end{figure}

\begin{figure}
  \centering
    \includegraphics[width=0.5\textwidth]{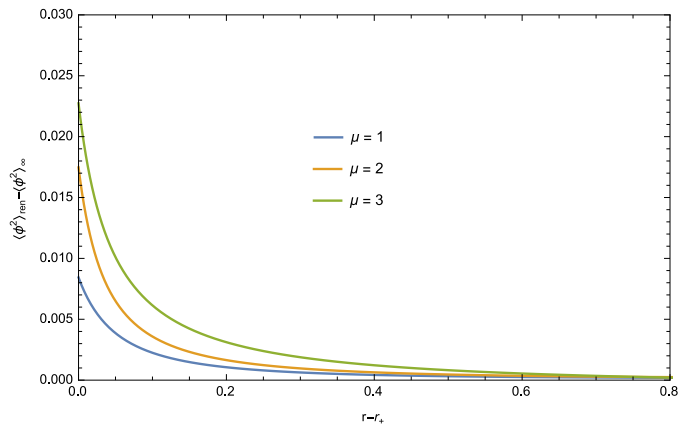}
      \caption{Plots of the vacuum polarization
        $\braket{\phi^2}_{\textrm{ren}}-\braket{\phi^2}_{\infty}$ as a
        function of the coordinate distance from the black hole
        horizon radius, i.e., $r-r_+$, for three scalar field masses
        $\mu$. The mass and charge of the black hole are fixed as $m =
        20$ and $q = 10$, respectively.}
\end{figure}

In Figs.~1-3, we plot
$\braket{\phi^2}_{\textrm{ren}}-\braket{\phi^2}_{\infty}$, i.e.,
the renormalized vacuum polarization normalized to zero at infinity, as a
function of the coordinate distance from the five-dimensional
Reissner-Nordstr\"om black hole horizon
radius, i.e., $r-r_+$, for three different values of the black hole
mass, black hole electric charge, and scalar field mass, respectively. For each
parameter choice, we find finite values at the horizon with no problems
of convergence.  Note that, since we deal with a five-dimensional
Reissner-Nordstr\"om black hole, the curvature $R$ is identically zero,
and so the coupling constant $\xi$ is irrelevant in our calculations.
In Fig.~1, we see that the value of the vacuum polarization at the horizon decreases with increasing black hole mass. This is expected, as the black hole temperature decreases and so it is harder to produce excitations in the quantum field.
In Fig.~2, the value at the horizon decreases with increasing charge, i.e., as the black hole approaches the extremal limit. This is again expected, as an extremal black hole has zero temperature. In Fig.~3, we see that increasing scalar field mass induces a larger vacuum polarization at the horizon.

\section{Conclusions}
\label{Conc}

In this work we have extended our previous results
\cite{Flachi:2016bwr} and calculated the renormalized vacuum
polarization for a massive scalar field around a five-dimensional
electrically charged black hole. We have followed the standard
approach which makes use of the WKB approximation to extract the
infinities present both in the angular and energy modes of the
mode-sum expanded Green function. We have also compared the explicit
divergent part with the Schwinger-DeWitt counterterms to get a fully
renormalized result for the vacuum polarization. Terms up to second
order were used in the approximation, which provided numerical results
illustrating the behavior of the vacuum polarization as a function of
the various parameters. A simple understanding of the finer features of the vacuum
polarization $\braket{\phi^2}_{{\rm ren}}$ in the various cases is
difficult due to the complexity of the calculations involved.

\section{Acknowledgments}
G.Q. acknowledges Funda\c c\~ao para a Ci\^encia e Tecnologia(FCT),
(Portugal) through Grant No.~SFRH/BD/92583/2013.  A.F. acknowledges the
MEXT-supported Program, Japan, for the Strategic Research Foundation
at Private Universities ``Topological Science,'' Grant No.~S1511006.
J.P.S.L. acknowledges FCT for financial support through
Project~No.~UID/FIS/00099/2013 and Grant No.~SFRH/BSAB/128455/2017,
and Coordena\c c\~ao de Aperfei\c coamento do Pessoal de N\'\i vel
Superior (CAPES), Brazil, for support within the Programa CSF-PVE,
Grant No.~88887.068694/2014-00.

\newpage
\end{document}